\documentclass[12pt]{elsart}
\usepackage{amssymb}
\begin{document}
\begin{frontmatter}

\title{Accelerated recombination due to resonant deexcitation of metastable 
states}

\author{Yuri V. Ralchenko} and
\author{Yitzhak Maron}

\address{Faculty of Physics, Weizmann Institute of Science, Rehovot 76100,
Israel}

\begin{abstract}
In a recombining plasma the metastable states are known to accumulate
population thereby slowing down the recombination process. We show that an
account of the doubly-excited autoionizing states, formed due to collisional
recombination of metastable ions, results in a significant acceleration of
recombination. A fully time-dependent collisional-radiative (CR) modeling
for stripped ions of carbon recombining in a cold dense plasma demonstrates
an order of magnitude faster recombination of He-like ions. The CR model
used in calculations is discussed in detail.
\end{abstract}

\end{frontmatter}

\section{\protect\bigskip Introduction}

Recombination of atomic ions in plasmas \cite{ReBook} continues to be a
subject of a permanent interest. Recent measurements of radiative
recombination cross sections for various ions, including the simplest bare
ions (see, e.g., Ref. \cite{Muller99} and references therein), have shown a
noticeable and surprising disagreement between theory and experiment. In
addition to such fundamental issues, various phenomena related to
recombination play important roles in plasma kinetics \cite{Griem97}. A
well-known example is provided by a recombination-based formation of
population inversion used for practical lasing in
the soft X-ray region \cite{Elton}.

In spite of a widely recognized importance of ion recombination for 
plasma evolution, the
theoretical efforts were hitherto directed practically only to the
recombinational kinetics of (quasi-)hydrogenic ions while non-hydrogenic
recombination was only scarcely studied \cite{Cacci76,Gorse78}. As far
as recombination is concerned, the major difference between hydrogenic and
non-hydrogenic ions is the possible presence of long-lived metastable states
in the energy spectrum of the latter. The metastable states accumulate the
downward flowing population for times of the order of the inverse
depopulation rates. Since this rate could be rather small, the population
accumulation would therefore exist for relatively long times. 
The purpose of this
paper is to demonstrate that the process of the so-called ''resonant
deexcitation'' (RD) can significantly alter this picture and, in fact,
accelerate the overall recombination process by as much as an order of
magnitude.

The resonant deexcitation, as it is defined here, proceeds in three steps.
First, a free electron is collisionally captured by an {\em excited} ion
with the spectroscopic charge $Z+1$ to form a doubly-excited state of the
ion $Z$. This step is a familiar collisional 3-body recombination which
though originates from excited rather than ground state of an ion. In the
second step, the captured electron may move upward or downward 
between 
high-$n$ states due to collisional excitation or deexcitation. Finally, the
doubly-excited state decays via autoionization thereby producing an ion with
the charge $Z+1$, so that the initial and final states of RD belong to the
same ion stage. The total chain of elementary events, i.e., recombination,
(de)excitation and autoionization, looks as follows:

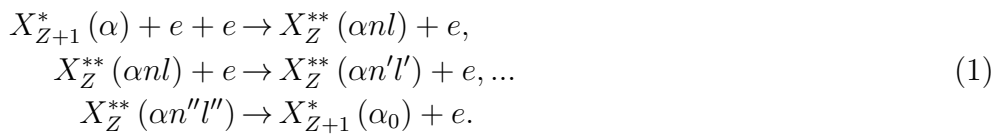
\begin{eqnarray}
X_{Z+1}^{\ast }\left( \alpha \right) +e+e &\rightarrow &X_{Z}^{\ast \ast
}\left( \alpha nl\right) +e,  \nonumber \\
X_{Z}^{\ast \ast }\left( \alpha nl\right) +e &\rightarrow &X_{Z}^{\ast \ast
}\left( \alpha n^{\prime }l^{\prime }\right) +e,...  \label{rd0} \\
X_{Z}^{\ast \ast }\left( \alpha n^{\prime \prime }l^{\prime \prime }\right) 
&\rightarrow &X_{Z+1}^{\ast }\left( \alpha _{0}\right) +e.  \nonumber
\end{eqnarray}
Here $\alpha $ denotes the quantum state of the initial excited ion, $nl$
are the principal and orbital quantum numbers of the captured electron, and $%
\alpha _{0}$ is the set of the final quantum numbers. The final state $%
X_{Z+1}^{\ast }\left( \alpha _{0}\right) $ may obviously be also an excited
one (rather than only the ground state) provided the energy difference between
initial and final states of the $Z+1$ ion is larger than the ionization
energy of the $n^{\prime \prime }l^{\prime \prime }$ electron. The
doubly-excited quantum states $X_{Z}^{\ast \ast }\left( \alpha nl\right) $\
with different $nl$\ form a shifted Rydberg series with the ionization
limit $X_{Z+1}^{\ast }\left( \alpha \right) $.

The effect of RD on level populations was independently recognized by
Fujimoto and Kato \cite{KaFu82} and by Jacoby {\em et al.} \cite{Jacoby}
(with respect to the X-ray laser problem). Later and independently, Koshelev
with colleagues \cite{Kosh89,Kosh92} discussed the population of
doubly-excited levels via 3-body recombination from excited states. The
primary interest in these papers was directed towards either hydrogen-like
ions \cite{KaFu82,Jacoby} or production of dielectronic satellites \cite
{Kosh89,Kosh92}, and therefore the importance of RD for recombination of
non-hydrogenic ions was not investigated. Moreover, it was found \cite{Jacoby}
that the RD channel 
\begin{equation}
n=3\rightarrow 3l\/nl^{\prime }\rightarrow 1s+e
\end{equation}
in the recombining H-like C {\sc VI} plasma is negligibly small
 comparing to the
direct radiative decay $n=3\rightarrow n=1$. Then, in a recent series of papers
Kawachi {\em et} {\em al.} \cite{Kawachi} investigated a quasi-steady-state
recombination of Li-like Al and showed that accounting for RD results in a
better agreement with experimental data. It should be noted that some
groups \cite{KaFu82,Kawachi} refer to the described above process of Eq. (\ref
{rd0}) as the ''DL-deexcitation'' which originates from the inverse process
of dielectronic capture and ladder-like excitation. Here the term ''resonant
deexcitation'' is preferred since no dielectronic capture is involved in the
process considered.

The paper is organized as follows. Section II contains a detailed
description of the collisional-radiative model used in the simulations. We
describe the major kinetic processes taken into account and indicate the
sources of atomic data. The calculational results for time-dependent 
recombination of carbon nuclei and discussion are
presented in Section III. Finally, Section IV contains conclusions.

\section{Collisional-Radiative Model}

The experimental installations at the Plasma Laboratory of the Weizmann
Institute of Science (coaxial and planar plasma opening switches and
Z-pinch) produce plasmas with diverse and fast-changing properties. The
computational tools required for reliable diagnostics of such systems should
therefore both reflect a variety of physical processes happening in plasmas
and consistently describe the temporal behavior of plasma characteristics
under very different conditions. The collisional-radiative (CR) package
NOMAD was developed to provide reliable spectroscopic diagnostics of
transient plasmas with an arbitrary electron energy distribution function.
It includes an executable code and a number of atomic databases containing
various related data for many elements and their ions. A convenient
user-friendly interface relies upon the usage of common Web browsers and
provides both textual and graphical options for the processing of calculated
results. The program was written in Fortran 77 with minor extensions from
Fortran 90. Presently, the code is running on a 450 MHz Pentium 
III personal computer with typical run times of the order of several
minutes.

Generally, the CR code solves the following first-order system of
inhomogeneous differential equations:

\begin{eqnarray}
\frac{d\hat{N}\left( t\right) }{dt}=\hat{A}\left( N_{i},N_{e},f_{e},t\right) 
\hat{N}\left( t\right) +\hat{S}(t), \nonumber\\
\hat{N}\left( t=0\right) =%
\hat{N}_{0},  \label{rateq}
\end{eqnarray}
where $\hat{N}(t)$ is the vector of atomic state populations, $\hat{A}\left(
N_{i},N_{e},f_{e},t\right) $ is the rate matrix depending on ion density $%
N_{i}(t)$, electron density $N_{e}(t)$ and electron-energy distribution
function $f_{e}(E,t)$, and $\hat{S}(t)$ is the source function. The electron
density can be presented as a sum of two components:

\begin{equation}
N_{e}\left( t\right) =N_{e}^{0}\left( t\right) +N_{i}\left( t\right)
\sum_{Z=Z_{min}}^{Z_{max}}\left( Z-1\right)
\sum_{k=1}^{k_{max}}N_{Z,k}\left( t\right) ,  \label{eden}
\end{equation}
where $N_{e}^{0}\left( t\right) $ is the background electron density and the
second term represents the density of continuum electrons originating from
ionization of atoms. In Eq. (\ref{eden}) $Z_{min}$ and $Z_{max}$ are the
minimal and maximal spectroscopic charges, index $k$ enumerates the levels
within a specific ion with charge $Z$ ($k=1$ is the ground state etc.), and $%
N_{Z,k}(t)$ is the population of the corresponding atomic level. 
The level populations are usually normalized:

\begin{equation}
\sum_{Z=Z_{min}}^{Z_{max}}\sum_{k=1}^{k_{max}}N_{Z,k}\left( t\right) =1, 
\end{equation}

although for calculations with the source function $\hat{S}(t)$ this condition
should be discarded.
The number of atomic states used in specific calculations can be chosen
depending upon the complexity of a task. There are no fundamental
limitations on the nature of the states involved, and therefore atomic
terms, levels (fine structure components) or configurations can be equally
used in the CR calculations if necessary\footnote{%
Below we use the terms ''state'' and ''level'' interchangeably.}. In addition,
an arbitrary number of high-$n$ Rydberg states can be added to each of the
ion charge states. The highest in energy bound state is determined by the
ionization potential lowering. This effect may be accounted for in different
approximations using Debye-Huckel, hybrid (Stewart-Pyatt) or ion-sphere
formulas \cite{Muri98}. However, an explicit account of numerous Rydberg
states essentially increases the calculation time and thus is not always
practical. In this case, an effective aggregate state, which is composed of
Rydberg states up to the highest bound state, may be used in calculations.

As was already mentioned above, the model allows for use of an arbitrary
electron-energy distribution function (EEDF). Usually, the EEDF is first to
have been calculated with a coupled hydrodynamic plasma code and then
utilized for a detailed CR modeling. It is possible, nevertheless, to
approximate this very complicated problem by using simplified EEDFs, for
example, Maxwellian+beam or a two-Maxwellian distribution. To this end, the
EEDF $f\left( E\right) $ is presented as:

\begin{equation}
f\left( E\right) =\left( 1-\alpha \right) f_{M}\left( T_{e},E\right) +\alpha
f^{\prime }\left( E\right) ,
\end{equation}
where $f_{M}(T_{e},E)$ is the Maxwellian EEDF with an electron temperature $%
T_{e}$, $f^{\prime }(E)$ is either the beam or the second Maxwellian EEDF,
and $0\leq \alpha \leq 1$ is the weight of $f^{\prime }(E)$ in the total
EEDF. A modeling with an arbitrary EEDF poses a serious restriction on the
data used for the calculation of rates of atomic processes, namely, cross
sections rather than Maxwellian-averaged rate coefficients have to be used
in calculations.

The atomic processes which affect the level populations and are considered in
our model include spontaneous radiative decays, electron-impact collisional
processes (excitation, deexcitation and ionization), various types of ion
recombination (3-body, radiative and dielectronic recombination),
autoionization and dielectronic capture, atom-ion or  ion-ion charge exchange,
and laser photopumping. The plasma opacity effects are taken into account as
well. The atomic data (either raw or fitted with physically 
justified formulas) are stored in databases, one per element, and
the data accuracy is carefully evaluated.

The energy levels and radiative oscillator strengths are mainly collected
from available publications and online databases\footnote{%
A list of Internet atomic databases can be found in \cite{DBfAPP}.}.
Whenever these sources cannot provide a necessary level of accuracy and/or
completeness, the missing energies and oscillator strengths are calculated
with various atomic software, e.g., RCN/RCN2/RCG Hartree-Fock package \cite
{Cowan1981}, MZ 1/Z-expansion code \cite{SheVan}, or GRASP92
multiconfiguration Dirac-Fock code \cite{Grasp}. 

The electron-impact collisional data comprise a majority of atomic data
utilized in CR calculations. The existing data on cross sections are rather
incomplete and mostly cover excitations and ionizations from ground 
states. Hence, production of new cross sections becomes a
necessity if a detailed diagnostics is required. Our main source of
excitation data is the Coulomb-Born-exchange unitarized code ATOM \cite
{SheVan} which combines a high calculational speed with a good accuracy, 
especially for moderately- and highly-ionized
atoms. Previously we have shown that ATOM
excitation data for H- and Li-like ions very well agree with the cross
sections calculated by the more sophisticated convergent close-coupling
(CCC) method \cite{Fisher97,Fish97a}. Nevertheless, a use of more precise
data is made whenever possible; for example, our database of collisional
cross sections for neutral Helium was developed using the latest recommended
CCC fits \cite{Ral00}. For comparison purposes or when no other
data are available, a simple van Regemorter formula for excitation cross
sections may be utilized as well. The inverse, deexcitation cross sections
are obtained using the detailed balance principle, the former being
implemented also for calculation of other inverse cross sections from 
the direct
ones (e.g., 3-body recombination from ionization etc.).

The fits for the electron-impact ionization cross sections from ground states
were obtained using the recommended data compiled by the Belfast group \cite
{Bell83,Lennon88} with account of corrections discussed in Ref. \cite
{KatoION}. Presently, four options for calculation of the
ionization cross sections from excited states are available, namely: 
(i) ATOM data, (ii)
Lotz formula \cite{Lotz}, (iii) $l$-dependent semiempirical formula \cite
{Bern00}, and (iv) simple classical scaling

\begin{equation}
\sigma (E/I)=\frac{I_{0}^{2}}{I^{2}}\sigma _{0}(E/I_{0})  \label{ion_cs}
\end{equation}
with $I$ and $I_{0}$ being the corresponding ionization energies from
excited and ground states, respectively. The single ionization is allowed to
proceed into both ground and excited states of the next ion. The multiple
ionization which may be very important for non-equilibrium plasmas,
especially in the presence of high-energy electrons, is also taken into
account. The corresponding cross section connecting ground states of
respective ions is calculated with recommended formulas from Ref. \cite
{Fisher}.

The photoionization cross sections, which are also used for calculation of
radiative recombination data with the
Milne formula, are compiled from a few sources, such as Opacity Project 
\cite{Opacity}, ATOM calculations or other published compilations and
evaluations \cite{photo,photo2}. The less accurate hydrogenic Kramers
formulas are used to calculate the photorecombination into the high-$n$
Rydberg states and/or the aggregate state.

The dielectronic recombination can be taken into account in two ways. One
method consists in an explicit treatment of autoionizing states with a use
of dielectronic capture cross sections and autoionization transition
probabilities. The latter are usually calculated with the MZ or Cowan's
\cite{Cowan1981} codes and the former are obtained using the principle of 
detailed balance.
Furthermore, the collisional and radiative transitions to, from, and between
all doubly-excited autoionizing states are fully accounted for. In the
second, less detailed approach, no autoionizing states are presented in the
CR calculations, and the dielectronic recombination rates are calculated
using one of the existing methods, namely, the modified Burgess formula 
\cite{Cowan1981}, fitting formulas of Hahn \cite{Hahn93}, or recent
recommended fits from Ref. \cite{Mazzotta}.

A two-step procedure is implemented in treatment of the plasma opacity
effects. First, when calculating the level populations, a simple escape
factor method is applied. The dependence of the escape factor on optical depth
is calculated using (i) the fitting formulas produced for a general Voigt line
profile with the Monte-Carlo code TRACE \cite{Schultz} or (ii) 
Apruzese's formulas for the Voigt escape factor \cite{Apruzese}. Then, on the
second step, when an actual spectrum is synthesized, the radiation transfer
equation for a Voigt profile is solved for each of the selected spectral
lines using the level populations obtained from the rate equations. The
Voigt line profile is constructed from Doppler (Gaussian) and natural+Stark
(Lorentz) broadening parts. The Doppler width is determined by the ion
temperature $T_{i}$ which may be another independent input parameter. The
depopulating rates from both lower and upper levels of specific radiative
transitions are used to calculate the inelastic Stark width, and the elastic
Stark linewidth is ignored because of its steep decrease with electron
temperature and small contribution for moderate to high temperatures (see
Refs. \cite{Stark1,Stark2} for discussion on elastic contribution). Finally,
it is worth noting that Drayson's routine \cite{Drayson} is used in the
generation of Voigt profile.

\section{Results and discussion}

To study the effect of resonant deexcitation on kinetics of recombination,
we consider here the time-dependent recombination of fully stripped ions of
carbon in an optically thin cold dense plasma. This situation may be
experimentally implemented, for instance, when a beam of bare ions is
injected into a pre-formed plasma. The recombination of carbon nuclei has
already been a subject of discussion at the NLTE kinetics workshop \cite
{NLTE97}; however, emphasis was given to the most general plasma
characteristics such as mean ion charge and therefore no detailed
examination of the evolution of level populations was carried out.

In the present calculation all charge states of carbon from neutral atom up
to the fully stripped ion were retained. The basic atomic states were mainly
the atomic terms characterized by the total angular momentum and spin. The
exceptions are the following: (i) the $1s2l$ $^{3}P$ term in C V is split
into the fine structure components, (ii) the doubly-excited autoionizing
states in Li-like C IV and He-like C V are represented by the configurations 
$1s2lnl^{\prime }$ (total of 15 states up to $1s2s8l$ and $1s2p8l$) and $%
2lnl^{\prime }$ (7 states up to $n=8$), and (iii) $l$-summed states
characterized only by the principal quantum number are used in H-like C VI.
Besides, 20 high-$n$ Rydberg states were added to each of the ion charge
states. However, the actual number of Rydberg states becomes smaller since
the ionization potential lowering effectively cuts off the bound spectrum.
The total number of atomic states was about 180.

The CR calculations were performed for two cases differing in the number
(and nature) of included states. In the first case ({\sf A}), 
the doubly-excited autoionizing states for C IV and C V 
are excluded from consideration, and thus the resonant deexcitation 
channel is closed. 
Nevertheless, we do account for the process
of dielectronic recombination using the rates recommended by Hahn \cite
{Hahn93}. It is worth mentioning that this type of ion recombination is
essentially unimportant for the low temperatures and high densities specific
for the problem in question. In the second case ({\sf B}), 22 doubly-excited 
states in C
IV and C V listed above were added with a detailed account of all relevant
atomic processes (autoionization, dielectronic capture, ionization, 3-body
recombination, radiative decays, and excitation/deexcitation). The electron
impact excitation and deexcitation cross sections between these levels were
calculated in the van Regemorter approximation, while the relevant radiative
and autoionization probabilities were determined with the {\rm MZ} code. In
addition, the collisional and radiative (satellite) transitions of the core
electron, e.g., $1s2pnl\rightarrow 1s^{2}nl+h\nu $, were also taken into
account.

For each of the cases {\sf A} and {\sf B}, the CR simulations were carried
out for several sets of electron density $N_{e}$ and temperature $T_{e}$.
Below we mainly discuss the plasma evolution for $N_{e}=3\times 10^{19}$ cm$%
^{-3}$ and $T_{e}=3$ eV, and comparison with other sets is given when 
necessary\footnote{%
This particular set of plasma parameters will be referred to as the ''basic
conditions''.}. Both electron density and temperature were kept constant
during the run. The initial population distribution and temporal history
were set as following. At time $t=0$ all population is in the bare nucleus of
carbon, and the logarithmic time mesh for basic conditions is chosen
according to:

\begin{eqnarray}
t_{0} &=&0; t_{1}=10^{-16} s, \nonumber \\
i &>&1: t_{i}=t_{i-1}\ast 1.11344,
\end{eqnarray}
so that $t_{150}=10^{-9}$ s. This final time was found to be sufficient to
reach the CR equilibrium (CRE) state for both sets of calculations ({\sf A}
and {\sf B}). For other than basic conditions, the logarithmic time mesh was
properly adjusted in order to achieve CRE within 150 steps.

Consider first the calculated evolution of level populations for case {\sf A}
(no doubly-excited autoionizing states). The total populations of
all charge states of carbon as a function of time are shown by the solid
lines in Fig. 1. The mostly noticeable feature in this plot is a very long
lifetime of the He-like C V ion, $t\simeq $ $10^{-10}$ s, which exceeds
the lifetime of C IV. To examine how the metastable states $2^{3}S$\ and $%
2^{3}P_{0..2}$ affect the C V lifetime, let us compare the populations of
all $n=1$\ and $n=2$\ states of C V and the total C V population (Fig.
2(a)). Note that the components of the $2^{3}P$ term are in local
thermodynamic equilibrium (LTE) due to high collisional rates. The sum of
populations of the triplet states $N(2^{3}S)+%
\sum_{j=0..2}N(2^{3}P_{j})$ shown by the solid line with squares nearly
coincides with the total C V population after $t\simeq 1.5\times 10^{-11}$ s
which indicates that the lifetime of C V is indeed governed by the $n=2$
triplet levels. The radiative decay of the triplet levels
to the ground state is negligibly small comparing to the collisional 
deexcitation
rates which for basic conditions are $2.3\times 10^{9}$ s$^{-1}
$ for $2^{3}S$ and $1.9\times 10^{10}$ s$^{-1}$ for the $2^{3}P$ term. These
values are by factor $3.3$ and $2.5$, respectively, smaller than the highest
collisional rates between the $n=2$ triplet and singlet levels:
$7.8\times 10^{9}$ s$^{-1}$ for $2^{3}S\rightarrow 2^{1}S$ and $4.9\times
10^{10}$ s$^{-1}$ for $2^{3}P\rightarrow 2^{1}P$. 
This, together with the high
collisional rates within the triplet and singlet subsystems and strong
radiative decay $2^{1}P\rightarrow 1^{1}S$ with the probability $A\approx
8.9\times 10^{11}$ s$^{-1}$, shows that the characteristic
depopulation time of the triplet levels is mainly determined by the
triplet-singlet collisional transitions followed by a fast radiative decay
of the $2^{1}P$ state.

The long lifetime of C V results in a relatively low peak populations of the
C IV and C III ions (see Fig. 1). As a result, both C V and C II
simultaneously have high populations during a long time of the order of $%
10^{-10}$ s. Had such a picture of coexisting ion states with very different
charges be true, it would allow for a new scheme of laser photopumping when,
e.g., the photons from the $2^{3}S-2^{3}P$ transition in a long-living
He-like ion would pump a transition with the same wavelength in another
low-charge ion, thereby giving rise to a possible population inversion.
However, the detailed calculations for the case {\sf B} disallow such a
possibility.

The case {\sf B} calculations which were performed with an explicit account
of autoionizing states show a remarkably different temporal behavior of the
total ion populations (dashed lines in Fig. 1). Recalling that the
doubly-excited autoionizing states were added to both He- and Li-like ions,
one may notice the earlier appearance of the He-like ion comparing to
case {\sf A}. However, the lifetime of the H-like ion decreases only by less
than 30\% due to the RD channel $n=2\rightarrow n=1$. This indicates that
for the basic conditions the resonant deexcitation of the $n=2$
state of C VI can hardly compete with its strong radiative decay to
the ground $n = 1$ state. The temporal history of the He-like and other ion 
stages
is, on the other hand, drastically different from the case {\sf A}
calculations. As is seen from Fig. 1, the lifetime of C V in case {\sf B} is
an order of magnitude smaller due to a much faster decay of the metastable
states resulting from the resonant deexcitation. The effect of a strong
depopulating RD channel can be clearly seen in Fig. 2(b), where the
populations of $n=1$ and all $n=2$ states for case {\sf B} are presented.
One can notice that both $2^{3}S$ and $2^{3}P_{j}$ states have now
populations smaller than that of the ground state while in case {\sf A}
their populations exceed that of the ground state by about two orders of
magnitude. Furthermore, due to a much larger population flux from C V, the
peak populations of C IV and C III are now significantly larger (Fig. 1). 
Thus, one
can see that account of resonant deexcitation of the metastable levels in C
V leads to drastic changes in evolution of different ions.

The physical picture of resonant deexcitation is quite straightforward.
The highest-$n$ states are quickly populated from
the metastable states due to high 3-body recombination rates. The radiative
decay of the core electron $2p\rightarrow 1s$ does not depend significantly
on $n$ as long as $n$ is sufficiently high. The rates of autoionization $1s2l
$\/$nl^{\prime }\rightarrow 1s^{2}+\varepsilon l^{\prime \prime }$ are lower
for higher principal quantum numbers ($A_{a}(n)\sim n^{-3}$), and so are the
radiative transition probabilities of the outer electron. For dense
low-temperature plasmas the main channel of depopulation for the highest-$n$
states is a ladder-like collisional deexcitation to lower doubly-excited
states. When the population flow reaches those states for which the
autoionization probability is comparable or larger than the deexcitation
rate, the downward flow is redirected into the C V ground state, so that the
lowest doubly-excited states have small populations.

An approximate formula for the RD rate can be found assuming that the
populations for the doubly-excited autoionizing states above the so-called
''thermal limit'' \cite{Kosh89} are in the Saha-Bolzmann equilibrium with
their ''parent'' excited state of the next ion \cite{Jacoby}. The thermal
limit is defined as the state for which the rate of collisional processes is
of the order of autoionization probability, and the corresponding principal
quantum number $n_{th}$ for the thermal limit may be estimated from the
expression \cite{Kosh89}:

\begin{equation}
n_{th}\simeq 300\times \frac{Z^{2/9}T^{1/9}}{N_{e}^{1/9}}p^{4/9},
\label{nth}
\end{equation}
where $p$ is the principal quantum number of the excited state of the
recombining ion. The value of $n_{th}$ is obviously a very weak function of
plasma parameters, and for the basic conditions and $p=2$ one has $%
n_{th}\simeq 4$. Thus, the RD rate can be determined from the following
equation:

\begin{eqnarray}
N_{Z+1}\left( \alpha \right) R_{RD}\left( \alpha -\alpha _{0}\right) 
&=&\sum_{n\geq n_{th}}^{n_{max}}N_{Z}\left( \alpha nl\right) A\left(
\alpha nl-\alpha _{0}\right)   \nonumber \\
&=&N_{Z+1}\left( \alpha \right) N_{e}\left( \frac{2\pi \hbar ^{2}}{mT_{e}}%
\right) ^{3/2}\frac{1}{2g\left( \alpha \right) }  \nonumber \\
&&\times \sum_{n\geq n_{th}}^{n_{max}}e^{Z^{2}Ry/n^{2}T_{e}}g\left( \alpha
nl\right) ~A\left( \alpha nl-\alpha _{0}\right)   \label{Rate}
\end{eqnarray}
where $N_{Z+1}\left( \alpha \right) $ is the population of the initial
(excited) state $\alpha $ with the statistical weight $g\left( \alpha
\right) $, $N_{Z}\left( \alpha nl\right) $ is the population of the
doubly-excited state formed by a capture of the continuum electron into
atomic level with quantum numbers $n$ and $l$ and $g\left( \alpha nl\right) $
is its statistical weight, and $A\left( \alpha nl-\alpha _{0}\right) $ is
the probability of the autoionization process $\left( Z,\alpha nl\right)
\rightarrow \left( Z+1,\alpha _{0}\right) +e$. The summation is extended
above the thermal limit up to the highest bound doubly-excited state with
the principal quantum number $n_{max}$ which is determined by the ionization
potential lowering. Here we assume that the ionization energies of the
autoionizing states are given by the hydrogen-like formula $%
I_{n}=Z^{2}Ry/n^{2}$ with $Ry$ = $13.61$ eV being the Rydberg energy. For
low temperatures, which are the subject of the present work, the first term in
this sum with $n=n_{th}$ gives an overwhelming contribution due to a strong
exponential dependence on $n$. In agreement with this conclusion our
modeling shows that the doubly-excited states $1s2l3l^{\prime }$ and $%
1s2l4l^{\prime }$ indeed provide the largest contribution to resonant
deexcitation. 

Retaining only the first term in Eq. (\ref{Rate}) and using a well-known 
expression for
autoionization probability in terms of the excitation cross section at 
threshold (see, e.g., \cite{SheVan}), we obtain the following formula for
the RD rate :

\begin{eqnarray}
R_{RD}\left( \alpha -\alpha _{0}\right)  &=&N_{e}\frac{g\left( \alpha
_{0}\right) }{g\left( \alpha \right) }\frac{4\bar{g}f_{\alpha _{0}\alpha }}{%
\sqrt{3}\Delta E_{\alpha _{0}\alpha }}\frac{Z^{2}Ry^{2}}{\hbar n_{th}^{3}}%
\left( \frac{2\pi \hbar ^{2}}{mT_{e}}\right) ^{3/2}\exp \left( \frac{Z^{2}Ry%
}{n_{th}^{2}T_{e}}\right)   \nonumber \\
&\approx &2\times 10^{-6}\frac{g\left( \alpha _{0}\right) }{g\left( \alpha
\right) }\frac{N_{e}(cm^{-3})}{n_{th}^{3}T_{e}^{3/2}(eV)}\exp \left( \frac{%
13.6\cdot Z^{2}}{n_{th}^{2}T_{e}(eV)}\right) .  \label{r2}
\end{eqnarray}
To derive the second equation, we use the values of $\Delta E_{\alpha
_{0}\alpha }=3/4\cdot Z^{2}Ry$ for the transition energy (for He- and H-like
ions), $f_{\alpha _{0}\alpha }=0.5$ for the corresponding oscillator
strength for the transition $\left( Z,\alpha _{0}\right) $ $-$ $\left(
Z,\alpha \right) $, and $\bar{g}=0.2$ for the threshold value of the Gaunt
factor. Using Eq. (\ref{nth}) for $n_{th}$, one has:

\begin{equation}
R_{RD}\left( \alpha -\alpha _{0}\right) \approx 2.9\times 10^{-14}\frac{%
g\left( \alpha _{0}\right) }{g\left( \alpha \right) }\frac{N_{e}^{4/3}}{%
Z^{2/3}T_{e}^{11/6}}\exp \left( 8\times 10^{-5}\frac{Z^{16/9}N_{e}^{2/9}}{%
T_{e}^{10/9}}\right) .  \label{rd2}
\end{equation}
Thus, the RD rate for low-temperature plasmas shows a strong dependence on
electron temperature (cf. $T_{e}^{-4.5}$ behavior of the 3-body
recombination rate) and a moderate, slightly stronger than linear,
dependence on electron density. The $T_e$-dependence is illustrated in 
Fig. 3 where
we present the time evolution of carbon charge states for $T_{e}=2,$ $5$ and 
$10$ eV and a constant density of $N_{e}=3\times 10^{19}$ cm$^{-3}$. It is
clearly seen that the difference between the simulations with (solid lines)
and without (dashed lines) RD is mostly noticeable for the lowest of
temperatures while for $T_{e}=10$ eV both sets of calculations produce
almost identical evolution of charge states.

The effective RD rate calculated from Eq. (\ref{rd2}) is about 
$6\times 10^{12}$ s$^{-1}$ for the basic conditions. This value seems 
to disagree
with the results of the detailed modeling (Fig. 1) for C V where the
characteristic decay time is of the order of $10^{-11}$ s. Such a
discrepancy may be in part due to strong deviations from the Saha-LTE regime
for the doubly-excited states during the fast recombination which is
considered here. Our calculations show that during the time when C V is 
mostly abundant, 
only
the highest autoionizing states $1s2l8l^{\prime }$ (and $1s2l7l^{\prime }$
to a lesser extent) have Saha populations while the others deviate strongly
from the respective Saha limits. As for the $1s2l3l^{\prime }$ and $%
1s2l4l^{\prime }$ states which mostly
contribute to resonant deexcitation, their populations differ by as much as
a factor of 5. Nevertheless, a use of the effective RD rate from
Eq. (\ref{rd2}) in CR calculations {\em without} doubly-excited states
results in charge state evolution
 which is rather close to that obtained from the detailed
simulations with autoionizing states and RD included (case {\sf B}). 
The charge state
populations shown in Fig. 4 were calculated for these two cases for $N_{e}=$ 
$10^{20}$ cm$^{-3}$ and $T_{e}=5$ eV. (Similar results are obtained for
other sets of plasma parameters.) One can see that both characteristic decay
times and peak ion populations are reproduced within a few tens of percent.
It is interesting to note that
this situation is similar to the coronal approximation where line intensities
do not depend on the radiative decay probability and are determined only by
the collisional excitation rates. In the present case, the RD rate is
equivalent to the radiative rate while the flux of population coming
from the upper states due to cascades is
analogous to excitation in corona. Thus, as long as the rate of resonant
deexcitation from metastable states exceeds other depopulation rates, the
population flux downward is determined only by the cascade contribution 
and thus is not too sensitive to the actual value of the RD rate.

\section{Conclusions}

Although resonant deexcitation is already known for a long time, its
significance for plasma kinetics does not seem to have been fully appreciated 
so far. 
A particular problem of
recombination of ions with metastable states shows that a consistent account
of resonant deexcitation via the doubly-excited autoionizing states, which
are collisionally populated from the excited states of the next ion, can
considerably alter the entire picture of recombination. This process may be
important, for instance, in kinetics of recombination lasers or for 
beam-stopping problems.
It would certainly be interesting to conduct experiments that could test the
conclusions made in the present paper.

\section{Acknowledgments}

The collaboration with V. Tsitrin on the early stages of this work is highly
appreciated. We are grateful to K. N. Koshelev for interesting discussions
and to H. R. Griem for valuable comments and reading of the manuscript.
Special thanks are due to V. I. Fisher for his help in development of the CR
model. The assistance of V. A. Bernshtam, A. Goldgirsh and A. Starobinets 
in development of atomic databases is highly appreciated. This work is
supported part by the Minerva foundation (Munich, Germany) and Israeli
Ministry of Absorption.

\newpage

{\bf Figure Captions}

Figure 1. Temporal history of carbon charge states for $N_{e}=3\times 10^{19}
$ cm$^{-3}$ and $T_{e}=3$ eV. Solid lines -- case {\sf A}, dashed lines --
case {\sf B}.

Figure 2. Time-dependent populations of the $1s^{2}$ and $1s2l$ states and C
V ion for $N_{e}=3\times 10^{19}$ cm$^{-3}$ and $T_{e}=3$ eV. (a) - case 
{\sf A}, (b) - case{\sf \ B}. The total population of $n = 2$ triplet
states is shown by a solid line with squares.

Figure 3. Temporal history of carbon charge states for $N_{e}=3\times 10^{19}
$ cm$^{-3}$ and $T_{e}=2,$ $5,$ and $10$ eV. Solid lines -- case {\sf A}, 
dashed lines -- case {\sf B}.

Figure 4. Temporal history of carbon charge states for $N_{e}=1\times 10^{20}
$ cm$^{-3}$ and $T_{e}=$ $5$ eV. Solid lines -- calculations with effective
resonant deexcitation rates, dashed lines -- full calculations with
autoionizing states included (case {\sf B}).


\begin{thebibliography}{99}
\bibitem{ReBook}  Recombination of Atomic Ions, ed. by W. G. Graham et al,
NATO ASI Series B: Physics Vol. 296 (Plenum Press, New York and London,
1992).

\bibitem{Muller99}  M\"{u}ller A. Phil Trans Royal Soc London 1999;
A357:1279-96.

\bibitem{Griem97}  Griem HR. Principles of Plasma Spectroscopy. Cambridge:
Cambridge Univ. Press, 1997.

\bibitem{Elton}  Elton RC. X-Ray Lasers. New York: Academic Press, 1990.

\bibitem{Cacci76}  Cacciatore M, Capitelli M. Z Naturforsch 1976; 31a:362-8.

\bibitem{Gorse78}  Gorse C, Cacciatore MA, Capitelli M. Z Naturforsch 1978;
33a:895-902.

\bibitem{KaFu82}  Fujimoto T, Kato T. Phys Rev Lett, 1982; 48:1022-25; Phys
Rev 1985; A32:1663-8.

\bibitem{Jacoby}  Jacoby D, Pert GJ, Shorrock LD, Tallents GJ. J Phys 1982;
B15:3557-80.

\bibitem{Kosh89}  Koshelev KN. J Phys 1989; B21:L593-6.

\bibitem{Kosh92}  Koshelev KN, Rosmej (Yartseva) ON, Rosmej FB, Hebach M,
Schulz A, Kunze H-J. J Phys 1992; B25:L243-7.

\bibitem{Kawachi}  Kawachi T, Fujimoto T. Phys Rev, 1997; E55:1836-42;
Kawachi T, Ando K, Fujikawa C, Oyama H, Yamaguchi N, Hara T, Aoyagi Y. J
Phys 1999; B32:553-62.

\bibitem{Muri98}  Murillo MS, Weisheit JC. Phys Rep 1998; 302:2-65.

\bibitem{DBfAPP}  URL http://plasma-gate.weizmann.ac.il/DBfAPP.html.

\bibitem{Cowan1981}  Cowan RD. The Theory of Atomic Structure and Spectra.
Berkeley: University of California Press, 1981.

\bibitem{SheVan}  Shevelko VP, Vainshtein LA. Atomic Physics for Hot
Plasmas. Bristol: IOP Publishing, 1993.

\bibitem{Grasp}  Parpia FA, Fischer CF, Grant IP. Comp Phys Comm 1996;
94:249-271.

\bibitem{Fisher97}  Fisher VI, Ralchenko YuV, Bernshtam VA, Goldgirsh A,
Maron Y, Golten H, Vainshtein LA, Bray I. Phys Rev 1997; A55:329-34; 

\bibitem{Fish97a}  Fisher VI, Ralchenko YuV, Bernshtam VA, Goldgirsh A,
Maron Y, Vainshtein LA, Bray I. Phys Rev 1997; A56:3726-33.

\bibitem{Ral00}  Ralchenko YuV, Janev RK, Kato T, Fursa DV, Bray I, de Heer
FJ. Report NIFS-DATA-59, 2000.

\bibitem{Bell83}  Bell KL, Gilbody HB, Hughes JG, Kingston AE, Smith FJ. J
Phys Chem Ref Data 1983; 12:891-916.

\bibitem{Lennon88}  Lennon MA, Bell KL, Gilbody HB, Hughes JG, Kingston AE,
Murray MJ, Smith FJ. J Phys Chem Ref Data 1988; 17:1285-1363.

\bibitem{KatoION}  Kato T, Masai K, Arnaud M. Report NIFS-DATA-14, 1991.

\bibitem{Lotz}  Lotz W. Z Phys 1967; 206:205-211.

\bibitem{Bern00}  Bernshtam VA, Ralchenko YuV, Maron Y. J Phys 2000;
B33:5025-32.

\bibitem{Fisher}  Fisher V, Ralchenko YuV, Goldgirsh A, Fisher D, Maron Y. J
Phys 1995; B28:3027-46.

\bibitem{Opacity}  URL http://vizier.u-strasbg.fr/OP.html.

\bibitem{photo}  Verner DA, Ferland GJ, Korista KT, Yakovlev DG. Astrophys J
1996; 465:487-98.

\bibitem{photo2}  Clark REH, Cowan RD, Bobrowicz FW. At Data Nucl Data
Tables 1986; 34:415-22.

\bibitem{Hahn93}  Hahn Y. JQSRT, 1993; 49:81-94.

\bibitem{Mazzotta}  Mazzotta P, Mazzitelli G, Colafrancesco S, Vittorio N.
Astr Astroph Suppl Ser 1998; 133:403-9.

\bibitem{Schultz}  Schulz A. PhD thesis, Ruhr-Universitat Bochum, 1990.

\bibitem{Apruzese}  Apruzese JP. JQSRT 1985; 34:447-452.

\bibitem{Stark1}  Griem HR, Ralchenko YuV, Bray I. Phys Rev 1999; E60:6241.

\bibitem{Stark2}  Ralchenko YuV, Griem HR, Bray I, Fursa DV. Phys Rev 1999;
A59:1890-5.

\bibitem{Drayson}  Drayson SR. JQSRT 1976; 16:611-7.

\bibitem{NLTE97}  Lee RW, Nash JK, Ralchenko Y. JQSRT 1997; 58:737-42.

\end{thebibliography}
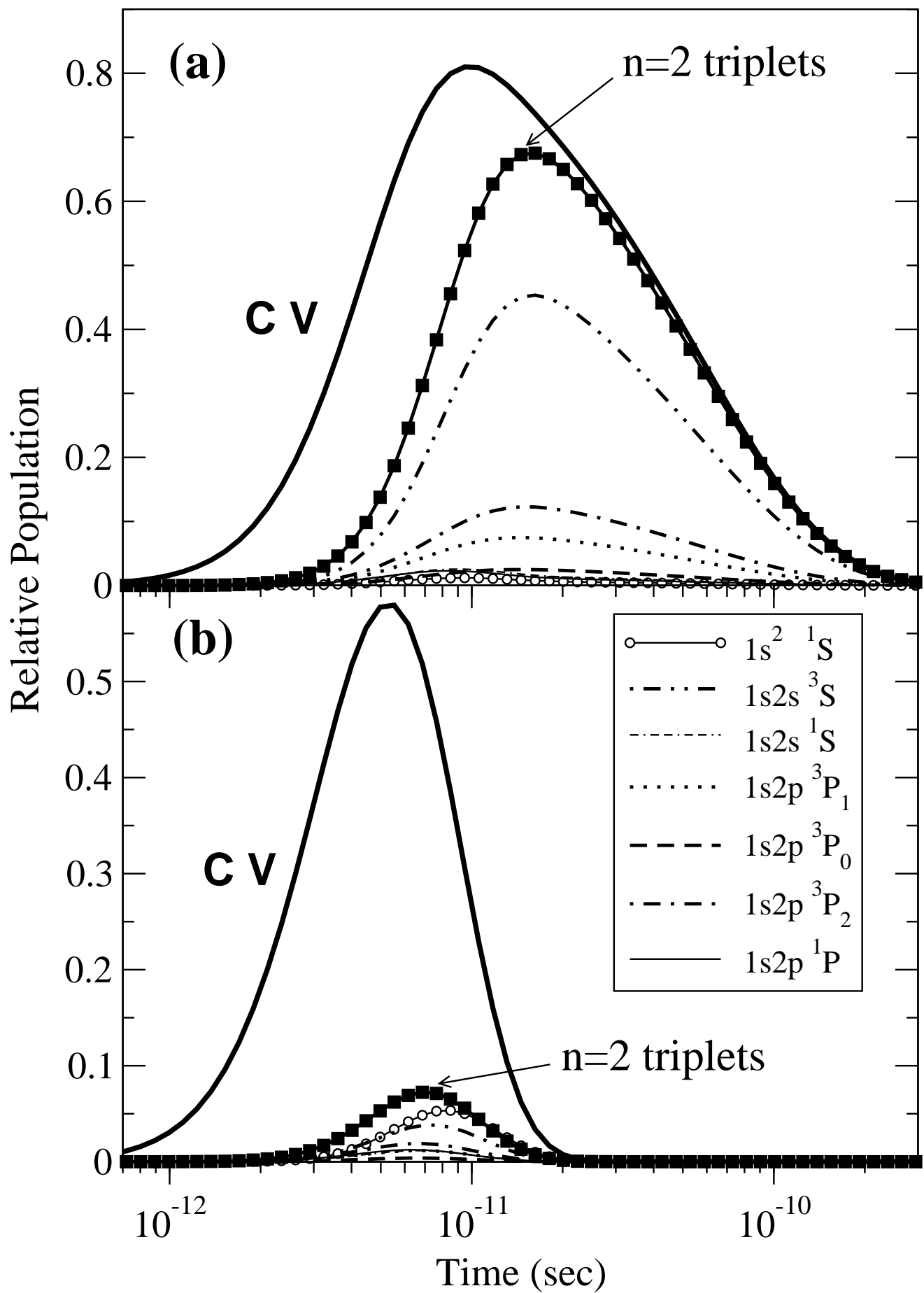
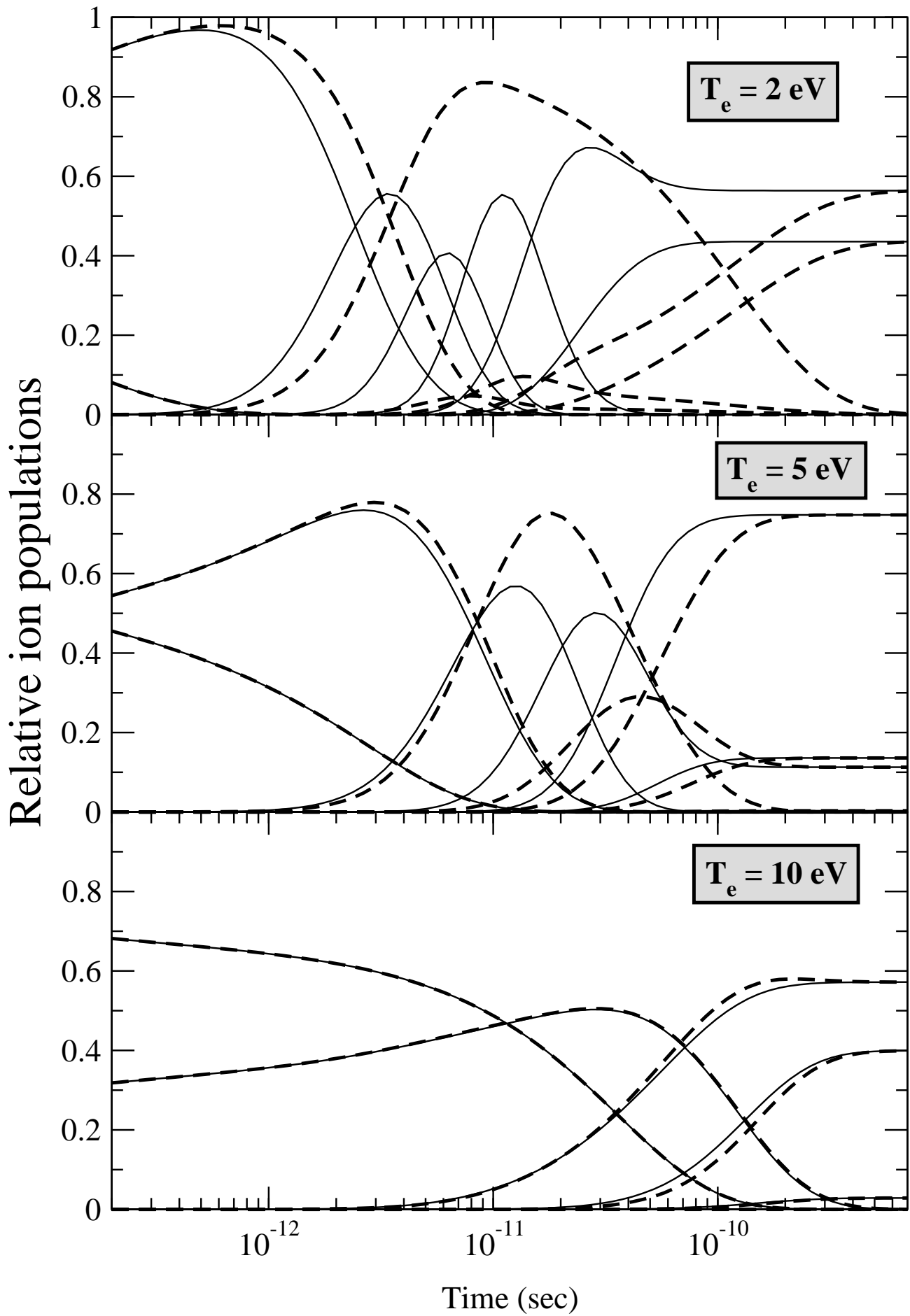
\end{document}